\documentclass[aps, prl, reprint, superscriptaddress]{revtex4-2}
\pdfoutput=1
\usepackage{braket,dsfont,subfigure,amsfonts,amssymb,bm,graphicx,float,amsmath,amsthm,txfonts}
\usepackage[colorlinks]{hyperref}
\usepackage{enumitem}
\usepackage{orcidlink}
\newcommand{\Tr}{{\rm Tr}}

\newcommand{\rrangle}{\rangle\!\rangle}
\newcommand{\llangle}{\langle\!\langle}

\newcommand{\llpipe}{|}
\newcommand{\LL}{\mathcal{L}}

\newcommand{\Z}{\mathbb{Z}}

\newcommand{\sket}[1]{\ensuremath{\llpipe#1\rrangle}}

\newcommand{\sbraket}[1]{\ensuremath{\llangle#1\rrangle}}
\theoremstyle{definition}

\begin{document}
\title{Strong-to-weak spontaneous symmetry breaking meets \\ average symmetry-protected topological order}
\author{Yuchen Guo~\orcidlink{0000-0002-4901-2737}}
\affiliation{State Key Laboratory of Low Dimensional Quantum Physics and Department of Physics, Tsinghua University, Beijing 100084, China}
\author{Shuo Yang~\orcidlink{0000-0001-9733-8566}}
\email{shuoyang@tsinghua.edu.cn}
\affiliation{State Key Laboratory of Low Dimensional Quantum Physics and Department of Physics, Tsinghua University, Beijing 100084, China}
\affiliation{Frontier Science Center for Quantum Information, Beijing 100084, China}
\affiliation{Hefei National Laboratory, Hefei 230088, China}

\begin{abstract}
    Recent studies have unveiled new possibilities for discovering intrinsic quantum phases that are unique to open systems, including phases with average symmetry-protected topological (ASPT) order and strong-to-weak spontaneous symmetry breaking (SWSSB) order in systems with global symmetry.
    In this work, we propose a new class of phases, termed the double ASPT phase, which emerges from a nontrivial extension of these two orders.
    This new phase is absent from prior studies and cannot exist in conventional closed systems.
    Using the recently developed imaginary-Lindbladian formalism, we explore the phase diagram of a one-dimensional open system with $\mathbb{Z}_2^{\sigma}\times \mathbb{Z}_2^{\tau}$ symmetry.
    We identify universal critical behaviors along each critical line and observe the emergence of an intermediate phase that completely breaks the $\mathbb{Z}_2^{\sigma}$ symmetry, leading to the formation of two triple points in the phase diagram.
    These two triple points are topologically distinct and connected by a domain-wall decoration duality map.
    Our results promote the establishment of a complete classification for quantum phases in open systems with various symmetry conditions. 
\end{abstract}

\maketitle

\emph{Introduction.\textemdash}
The exploration of many-body quantum matter in open quantum systems has recently attracted significant interest.
A key focus has been on the robustness of nontrivial quantum phases, such as the toric code with $\Z_2$ topological order~\cite{Levin2005, Kitaev2006, Levin2006, Levin2012}, against environmental coupling.
This has been examined from both the perspectives of topological properties, such as topological entanglement entropy and anyon condensation~\cite{Sang2024, Chen2024, Sang2025, Wang2024B}, and from the viewpoint of error correction, particularly the breakdown of quantum memory~\cite{Dennis2002, Wang2003, Fowler2012, Fan2024}.
Another compelling direction is the search for a complete classification of systems with intrinsic topological order (long-range entanglement)~\cite{Wang2023B, Sohal2024, Ellison2024}, as well as the symmetry protected topological (SPT) order~\cite{DeGroot2022, Ma2023A, Ma2023B, Guo2024A, Ma2024, Xue2024} or the spontaneous symmetry breaking (SSB) order~\cite{Lieu2020, Lee2022, Lee2023, Ogunnaike2023, Lessa2024B, Sala2024, Gu2024, Huang2024, Kuno2024} under global symmetry.

In contrast to the definition of symmetry $K$ for a pure state $\ket{\psi}$, where $K\ket{\psi} = e^{i\theta}\ket{\psi}$ and $\theta$ represents the corresponding symmetry charge, open systems possess two types of symmetry: strong symmetry, where $K\rho = e^{i\theta}\rho$, and weak symmetry, where $U\rho
U^{\dagger} = \rho$.
Consequently, open systems reveal new quantum phases that lack pure-state counterparts.
For instance, intrinsic average symmetry-protected topological (ASPT) order~\cite{Ma2023A, Ma2023B} is characterized by a nontrivial group extension of strong and weak symmetry and the mixed anomaly between them.
Another example is the strong-to-weak spontaneous symmetry breaking (SWSSB) order~\cite{Lessa2024B, Sala2024} characterized by the long-range pattern in terms of R\'enyi-2 or fidelity correlators.
These studies have constructed fixed-point density matrices for various quantum phases, though phase transitions between them remain unexplored because of the absence of a Hamiltonian and the definition of an energy gap in open systems.

On the other hand, in closed systems, a global symmetry can only protect a nontrivial SPT phase if it remains unbroken, either explicitly or spontaneously.
However, in open systems, a symmetry that experiences SWSSB can still protect an ASPT phase if supported by another strong symmetry.
This coexistence of SWSSB and ASPT within the same symmetry group is impossible in closed systems, revealing a new class of intrinsic mixed-state quantum phases that has not been previously identified in the study of open-system quantum phases.

To investigate phase transitions between different phases, a straightforward approach is to apply a noise channel to the quantum state, focusing on the decoherence-induced phase transition that typically occurs between a nontrivial phase and a trivial product phase~\cite{Chen2024, Sang2025, Wang2024B, Lessa2024B}.
A more comprehensive method is the imaginary-time Lindbladian formalism recently proposed ~\cite{Guo2024C, SeeSM}.
In this approach, a gapped quantum phase in open systems is defined by the finite-time steady state of the imaginary-time Lindbladian evolution, enabling the exploration of phase transitions between arbitrary two different phases provided with their representative states.
There remain several open questions in this framework, such as the rigorous connection between the imaginary-Liouville gap and the equivalent relation defined by two-way connectivity using local quantum channels.
We will demonstrate that, although without analytical proof, both numerical models in this study validate the effectiveness of this framework in terms of characterizing the universal properties of phase transitions in open systems.

In this paper, we focus on characterizing a nontrivial combination of SWSSB and ASPT phases in one-dimensional (1D) open systems.
We begin by investigating the phase transition between a trivial $\Z_2$ strong symmetric ($\Z_2$ (S)) phase and a $\Z_2$ SWSSB phase.
From the supervector perspective, this transition belongs to the conventional Ising criticality, although it exhibits distinct patterns for linear and R\'enyi-2 correlation functions of the corresponding density matrix in the SWSSB phase.
Building on this criticality, we propose a phase diagram for systems with $\Z_2^{\sigma} \times \Z_2^{\tau}$ symmetry.
Specifically, we introduce a new quantum phase, denoted as the double ASPT phase, which has no pure-state counterpart and has not been discovered in previous studies.
In this phase, the $\Z_2^{\sigma}$ (S) symmetry undergoes SWSSB, while the remaining $\Z_2^{\sigma}$ weak ($\Z_2^{\sigma}$ (W)) $\times$ $\Z_2^{\tau}$ (S) symmetry jointly protects a nontrivial ASPT phase.
Our phase diagram further demonstrates three types of SSB and three types of gapless ASPT phases, with two topologically distinct triple points intersected by different critical lines.
These results suggest more possibilities for constructing both quantum phases and quantum criticality that are intrinsic in open systems, especially with the presence of different symmetry types.

\emph{$\Z_2$ SWSSB.\textemdash}
As a preliminary step, we examine the phase transition between a trivial phase and the recently proposed SWSSB phase~\cite{Lessa2024B, Sala2024} for a single global $\Z_2$ symmetry.
The representative density matrices for these two phases are $\rho_0 = \prod_i\ket{\rightarrow_i}\hspace{-0.5mm}\bra{\rightarrow_i}$ and $\rho_1 \equiv O_+^{\rm SWSSB}+O_-^{\rm SWSSB} = \left( \frac{1}{2^N} \prod_i \sigma_i^0 \right) + \left(\frac{1}{2^N} \prod_i\sigma_i^x \right) $, respectively.
Both states exhibit the global $\Z_2$ (S) symmetry $K=\prod_i \sigma_i^x$, with $K\rho_i = \rho_i$ for $i=0, 1$.
In particular, $\rho_1$ represents a macroscopic superposition of two states in the double Hilbert space, where each state spontaneously breaks the $\Z_2$ (S) symmetry and only preserves a $\Z_2$ (W) symmetry, i.e., $KO_+^{\rm SWSSB}=O_-^{\rm SWSSB}$ and $KO_-^{\rm SWSSB}=O_+^{\rm SWSSB}$, but $KO_\alpha^{\rm SWSSB} K^{\dagger}=O_\alpha^{\rm SWSSB}$ for $\alpha=+, -$.

To study the phase transition between the two phases described above, we first construct the corresponding open systems for the fixed-point density matrices as follows~\cite{SeeSM}
\begin{align}
    &\textrm{Trivial: }H_0=-\sum_i\sigma_i^x, \quad L_0=0,\\
    &\textrm{SWSSB: }H_1=0, \quad L_{1, i}^{[1]}=\sqrt{2}\sigma_i^x,\quad L_{1, i}^{[2]}=\sqrt{2}\sigma_i^z\sigma_{i+1}^z.\label{Equ: SWSSB}
\end{align}
The imaginary-Liouville superoperator for each state can be subsequently defined as~\cite{Guo2024C}
\begin{align}
    \LL^I = \left(H_{\rm eff}^I\otimes I + I\otimes H_{\rm eff}^{I*}-\sum_k L_k \otimes L_k^{*}\right),
\end{align}
where $H_{\rm eff}^I = H - \sum_k \frac{1}{2}L_k^{\dagger}L_k$.
Finally, we obtain a parameterized model $\LL^I(\beta) = (1-\beta)\LL^I_0+\beta\LL^I_1$, which allows us to simulate the corresponding steady-state along this path and identify potential critical properties.
Notably, $\LL_i^I$ only contains commuting terms and hence it can be analytically verified that the above fixed-point density matrices $\rho_i$ are indeed the steady states of the corresponding $\LL_i^I$.
\begin{figure}
    \centering
    \includegraphics[width=\linewidth]{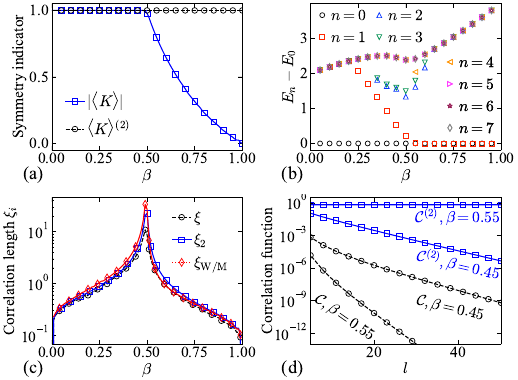}
    \caption{Phase transition between trivial phase and SWSSB phase.
    (a) Symmetry indicators for strong and weak symmetry for $O_+$.
    (b) Imaginary Liouville spectrum for $N=10$ under PBC.}
    (c) Linear, R\'enyi-2, Wightman, and Markov correlation lengths for $O_+$.
    (d) Linear and R\'enyi-2 correlation functions for $\rho=O_++O_-$.
    \label{Fig: SWSSB}
\end{figure}

In the following, we approximate the supervector $\sket{\rho}$ of the density matrix using a uniform matrix product state (MPS)~\cite{Perez2007, Orus2014, Haegeman2016, ZaunerStauber2018}, with a physical dimension $d_p=4$ (to account for both bra and ket spaces) and a virtual bond dimension $D=16$.
We simulate steady-states for different values of $\beta$ using the time-dependent variational principle (TDVP) method~\cite{Haegeman2011, Vanderstraeten2019}, a cutting-edge method that ensures high accuracy in characterizing a many-body system.
To investigate the nature of these steady states, we focus on different types of observables: linear, R\'enyi-2, and Wightman observables~\cite{SeeSM}.
These observables allow us to examine both standard expectation values and higher-order correlations within the system.
We also consider the associated correlation functions $\mathcal{C}, \mathcal{C}^{(2)}, \mathcal{C}^{\rm(W)}$, as well as conditional mutual information (CMI) and define their corresponding correlation lengths $\xi, \xi_2, \xi_{\rm W}\textrm{, and }\xi_{\rm M}$, providing a deeper understanding of phase transitions~\cite{Guo2024A, Lessa2024B, Liu2024, Sang2025}.

To maintain numerical stability, we introduce a small perturbing field $L_i^{\prime} = \sqrt{h}\sigma_i^z$ with $h=10^{-3}$ to explicitly break the strong symmetry to a weak one, leading to the steady state $O_{+}$.
In the trivial symmetric phase, $O_{+}$ approximately obeys the strong symmetry condition, while in the SWSSB phase, $O_+$ only maintains the weak symmetry.
Fig.~\ref{Fig: SWSSB}(a) compares two symmetry indicators $|\braket{K}|$ and $\braket{K}^{(2)}$~\cite{Guo2024C, SeeSM}, highlighting a phase transition at $\beta=0.5$ between the trivial and SWSSB phases.

To explore the nature of this phase transition, we plot the imaginary Liouville spectrum (the smallest eight eigenvalues) for a finite system with $N=10$ and periodic boundary condition (PBC) in Fig.~\ref{Fig: SWSSB}(b).
The results clearly show the process of gap closing when driven across the critical point from the trivial phase to the SWSSB phase, together with the two-fold degeneracy in the SWSSB phase.
The key signatures of this transition are also evident in the divergent behaviors of different correlation lengths of $O_+$ in Fig.~\ref{Fig: SWSSB}(c) and the entanglement entropy (EE) of the supervector 
$\sket{O_+}$ in Fig.~S3(a)~\cite{SeeSM}.
These observations not only validate the strong relation between the existence of a finite imaginary Liouville gap (if not degenerate) and the finite EE or correlation lengths, but also signals a transition characterized by an Ising-type criticality, where one of the $\Z_2$ symmetries is spontaneously broken in the double Hilbert space.
Therefore, our results validate the effectiveness of utilizing the properties of the imaginary-Liouville spectrum to define gapped or gapless phases as proposed in Ref.~\cite{Guo2024C}.

However, recent work points to a subtle distinction between SWSSB and the standard SSB observed in closed systems~\cite{Lessa2024B}.
Specifically, the SWSSB phase displays contrasting behavior between the linear correlator, which decays exponentially, and the R\'enyi-2 or Wightman correlator that remains nondecaying, as implied by the degeneracy of steady states in the imaginary Liouville spectrum.
In particular, any long-range order in the SSB phase is only captured by states that restore the broken symmetry, which explains why the correlation lengths $\xi_2$ and $\xi_{\rm W}$ remain finite in the SWSSB phase in Fig.~\ref{Fig: SWSSB}(c).
To recover the $\Z_2$ (S) symmetry, we construct $O_- = KO_+$ and compare the linear correlator $\mathcal{C}(\sigma^x, j, j+l)$ with the R\'enyi-2 correlator $\mathcal{C}^{(2)}(\sigma^z, j, j+l)$ in Fig.~\ref{Fig: SWSSB}(d) for $\rho=O_+ + O_-$.
In both phases, $\mathcal{C}$ exhibits exponential decay, while $\mathcal{C}^{(2)}$ saturates to a constant value in the SWSSB phase.
In this case, the Wightman correlator also shows a long-range pattern in the SWSSB phase (not shown here), though the concrete constant coefficient is difficult to determine~\cite{SeeSM}.
In summary, all correlators reveal divergent correlation lengths at the critical point, indicating a gapless density matrix, but only the R\'enyi-2 and Wightman correlators signal the presence of long-range order in the SWSSB phase.

\emph{SWSSB to ASPT.\textemdash}
Here, we explore an alternative scenario for a system exhibiting SWSSB, where the resulting state after symmetry breaking, such as $O_{+}$ in the previous example, belongs to a nontrivial ASPT phase.
In this case, the phase is protected by the interplay of the remaining weak symmetry and another strong symmetry that remains unbroken.
Unlike the previous example, which involves SWSSB to a trivial phase, this case represents SWSSB to an ASPT phase, characterized by a mixed anomaly between weak and strong symmetries~\cite{Ma2023A, Ma2023B, Guo2024A}.
This phase, termed double ASPT, represents a novel class of phases that has not been identified in earlier studies of SWSSB and falls outside the classification of ASPT phases, which generally assume short-range correlations for all correlation lengths.
The introduction of long-range correlations due to SWSSB challenges previous assumption and opens a pathway to discovering new quantum phenomena in open systems.

\begin{figure*}
    \centering
    \includegraphics[width=\linewidth]{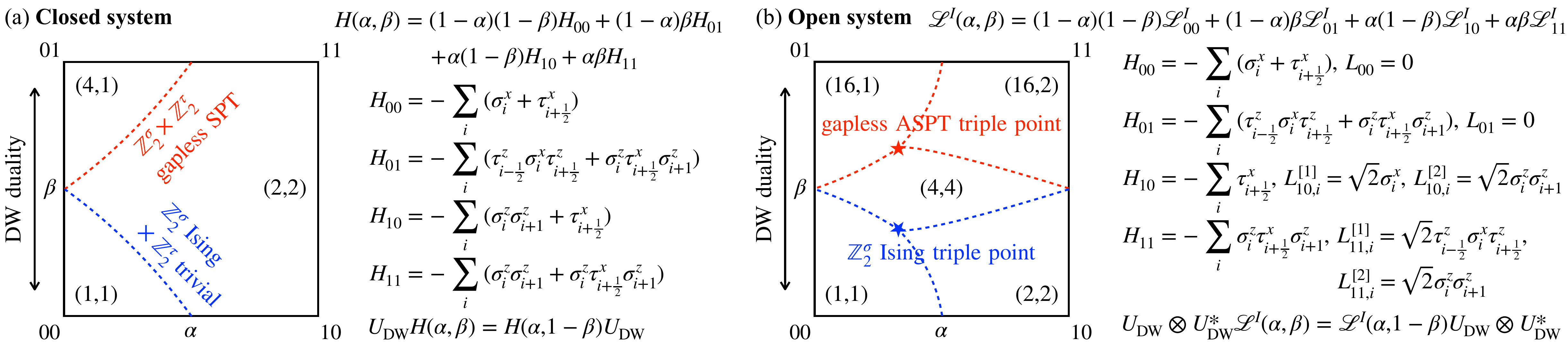}
    \caption{Phase diagram for a (a) closed system and (b) open system with $\Z_2^{\sigma} \times \Z_2^{\tau}$ symmetry.
    Each model is constructed by bilinear interpolation from four corner Hamiltonians or imaginary-Lindbladian superoperators.
    We mark the ground state degeneracy (GSD) under OBC and PBC by a tuple of two numbers under the label of each phase.}
    \label{Fig: Phase diagram}
\end{figure*}

To realize such a subtle pattern, a minimal symmetry group of $\Z_2\times \Z_2$ is necessary.
We begin with the pure cluster state, which belongs to a nontrivial SPT phase protected by $\Z_2^{\sigma}\times\Z_2^{\tau}$ symmetry commonly referred to as the Haldane phase~\cite{Pollmann2010, Pollmann2012, Chen2013}.
The wavefunction of this state can be expressed as
\begin{align}
    \ket{\psi} = \frac{1}{2^{N/2}}\sum_{\{\sigma_i\}}\ket{\cdots\uparrow_\sigma\rightarrow_{\tau}\uparrow_{\sigma}\rightarrow_{\tau}\uparrow_{\sigma}\leftarrow_{\tau}\downarrow_{\sigma}\leftarrow_{\tau}\uparrow_{\sigma}\cdots}\label{Equ: Cluster}.
\end{align}
There are two spins at each site ($\sigma_i$ and $\tau_i$), where the $\sigma$ spins form an equal-weight superposition and the excitations of the $\tau$ spins are positioned at the domain walls of the $\sigma$ spins.
The corresponding Hamiltonian for this cluster state is given by
\begin{align}
    H=-\sum_i\left(\tau_{i-1 / 2}^z \sigma_i^x \tau_{i+1 / 2}^z+ \sigma_i^z \tau_{i+1 / 2}^x\sigma_{i+1}^z\right).
\end{align}
If one introduces a ferromagnetic interaction $-\sigma_i^z\sigma_{i+1}^z$ and gradually increases its strength, the $\Z_2^{\sigma}$ symmetry will undergo SSB, driving the system into a $\Z_2^{\sigma}$ SSB $\times$ $\Z_2^{\tau}$ trivial phase since $\Z_2^{\tau}$ alone cannot support any nontrivial SPT phase.
The critical point of this transition is a gapless SPT phase, characterized by a gapless bulk and nontrivial topological edge modes~\cite{Scaffidi2017, Li2022, Li2024, Yu2024}.
Consequently, a possible phase diagram for a closed system with $\Z_2^{\sigma}\times\Z_2^{\tau}$ symmetry is shown in Fig.~\ref{Fig: Phase diagram}(a), where a parameterized Hamiltonian $H(\alpha, \beta)$ is constructed from a bilinear interpolation of four corner Hamiltonians, with $H_{11}$ and $H_{10}$ belonging to the same phase.
This model exhibits an intriguing duality named domain wall (DW) decoration map 
\begin{align}
    U_{\rm DW} \equiv \prod_{i}\textrm{CZ}_{i-1/2, i}^{\tau, \sigma}\textrm{CZ}_{i, i+1/2}^{\sigma, \tau}
\end{align}
inherited from the corner Hamiltonians, playing a crucial role in understanding the phase transitions within the system~\cite{Scaffidi2017}.

Let us now move on to the setting of an open system.
As discussed previously, starting from the cluster state and enforcing SWSSB on $\sigma$ spins while maintaining the decorated domain wall structure of $\tau$ spins leads to $O_+$ (after $\Z_2^{\sigma}$ SWSSB) that belongs to a nontrivial ASPT phase protected by $\Z_2^{\sigma}$ (W) $\times$ $\Z_2^{\tau}$ (S).
To derive a concrete model realizing this double ASPT phase, we extend the model used for SWSSB in Eq.~\eqref{Equ: SWSSB} by introducing an additional paramagnetic term $-\sum_i\tau_{i+1/2}^x$.
We then apply the generalized DW duality map~\cite{Guo2024C}
\begin{align}
    \LL^I_{\rm {Double\,ASPT}} = \left(U_{\rm{DW}}\otimes U_{\rm{DW}}^{*}\right)\LL^I_{\rm {SWSSB}}\left(U_{\rm{DW}}\otimes U_{\rm{DW}}^{*}\right)^{\dagger}
\end{align}
mapping the SWSSB phase into a double ASPT phase.
Compared to the pure-state case in Fig.~\ref{Fig: Phase diagram}(a), the phase diagram for the open system exhibited in Fig.~\ref{Fig: Phase diagram}(b), constructed through the bilinear interpolation of imaginary-Liouville superoperators, shows several notable differences.
\begin{enumerate}[leftmargin = 3mm, itemsep=-1mm, topsep=0mm]
    \item $\LL^I_{10}$ and $\LL^I_{11}$ belong to different phases.
    \item An intermediate phase where $\Z_2^{\sigma}$ (S) is completely broken (strong to none spontaneous symmetry breaking, or SNSSB) emerges as demonstrated later.
    \item There are two triple points, each defined by the intersection of three critical lines.
\end{enumerate}

\emph{Numerical results.\textemdash}
\begin{figure*}
    \centering
    \includegraphics[width=\linewidth]{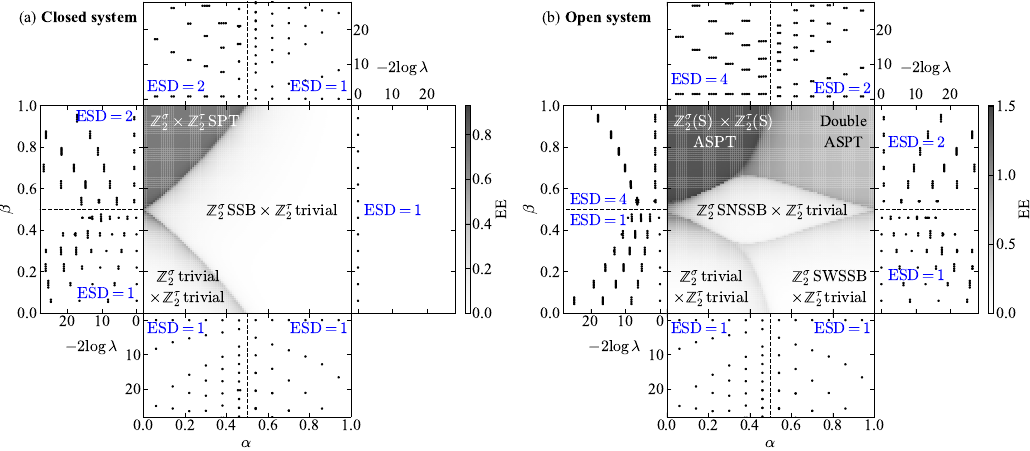}
    \caption{Entanglement entropy and entanglement spectrum along four axes for (a) closed system and (b) open system.
    We mark the entanglement spectrum degeneracy (ESD) for each phase in the lateral insets.}
    \label{Fig: SWSSB-ASPT}
\end{figure*}
In a similar approach to the previous example, an additional small field $L_i^{\prime} = \sqrt{h}\sigma_i^z$ with $h=10^{-3}$ is applied to explicitly break the $\Z_2^{\sigma}$ (S) symmetry.
The steady state is modeled using a uniform MPS with $d_p=16$ (to account for both $\sigma$ and $\tau$ spins on the ket and bra spaces) and $D=16$.
In Fig.~\ref{Fig: SWSSB-ASPT}, we plot the entanglement spectrum (ES) (of the wavefunction $\ket{\psi}$ for the pure state and of the supervector $\sket{O_+}$ for the mixed state) along four lines $\alpha=0$, $\alpha=1$, $\beta=0$, and $\beta=1$.
Additionally, we provide EE across the entire parameter space to offer a global view of the phase diagram. 
The entanglement spectrum degeneracy (ESD) is labeled with blue marks in each region.
To further characterize the system, the ground state degeneracy (GSD) under both open boundary conditions (OBC) and PBC is denoted in Fig.~\ref{Fig: Phase diagram}.
Other important observables, including various correlation lengths, two symmetry indicators, the string order parameter, and the imaginary Liouville spectrum, are shown in Fig.~S4-S6~\cite{SeeSM}, validating the classification of each phase in our phase diagram.

Overall, a comparison between Fig.~\ref{Fig: SWSSB-ASPT}(a) and \ref{Fig: SWSSB-ASPT}(b) shows three quantum phases in open systems also present in closed systems, namely $\Z_2^{\sigma}$ (S) $\times$ $\Z_2^{\tau}$ (S) ASPT, $\Z_2^{\sigma}$ trivial $\times$ $\Z_2^{\tau}$ trivial, and $\Z_2^{\sigma}$ SNSSB $\times $ $\Z_2^{\tau}$ trivial, where ESD or each GSD is the square of that in the corresponding wavefunction.
For instance, in the left panel ($\alpha=0$) of each diagram, the system undergoes a phase transition between the trivial and the SPT phase, where the ES of the supervector in the SPT phase is four-fold degenerate.
Interestingly, this topological phase remains stable even when coupled to the environment, provided the $\Z_2^{\sigma}$ (S) symmetry is preserved without any symmetry breaking.

In addition, the mixed-state phase diagram identifies two new phases absent from the pure-state system, namely $\Z_2^{\sigma}$ SWSSB $\times$ $\Z_2^{\tau}$ trivial and double ASPT phase.
In particular, each phase exhibits at least one ESD or GSD that is not a square of an integer, suggesting the lack of any pure-state counterpart.
In the $\Z_2^{\sigma}$ SWSSB $\times$ $\Z_2^{\tau}$ trivial phase, $O_+$ after symmetry breaking represents a trivial state with a nondegenerate ES.
However, $O_+$ in the double ASPT phase shows a two-fold degenerate ES, consistent with previous results for a single ASPT state protected by $\Z_2^{\sigma}$ (W) $\times$ $\Z_2^{\tau}$ (S) symmetry~\cite{Ma2023A, Ma2023B, Guo2024A}.

\begin{table}
    \caption{Number of edge modes, symmetry to be broken, and GSD under OBC and PBC for each phase.}
\begin{tabular}{c|c|c|c|c}\hline\hline
     Phase & Edge modes & SSB & $\rm{GSD_{OBC}}$ & $\rm{GSD_{PBC}}$ \\\hline
     $\Z_2^{\sigma}$ trivial $\times$ $\Z_2^{\tau}$ trivial & $1$ & $I$ & $1$ & $1$\\
     $\Z_2^{\sigma}$ SWSSB $\times$ $\Z_2^{\tau}$ trivial & $1$ & $\Z_2^{\sigma}$ & $2$ & $2$\\
     $\Z_2^{\sigma}$ SNSSB $\times $ $\Z_2^{\tau}$ trivial & $1$ & $\left(\Z_2^{\sigma}\right)^{\otimes 2}$ & $4$ & $4$\\
     $\Z_2^{\sigma}$ (S) $\times$ $\Z_2^{\tau}$ (S) ASPT & $2^4=16$ & $I$ & $16$ & $1$\\
     Double ASPT & $2^3=8$ & $\Z_2^{\sigma}$ & $16$ & $2$\\\hline\hline
    \end{tabular}
    \label{Tab: GSD}
\end{table}
GSD arising from symmetry breaking manifests itself under both OBC and PBC, while the degeneracy due to topological edge modes is only present under OBC. 
If a symmetry group $G$ undergoes SWSSB or SNSSB, the GSD follows the relation $\mathrm{GSD_{PBC}} = |G|$ or $\mathrm{GSD_{PBC}} = |G|^2$, respectively, where $|G|$ denotes the number of elements in $G$.
Under OBC, the topological edge modes introduce additional degeneracy, which depends on the symmetry group that protects these modes.
For example, in the double ASPT phase, three independent dangling spins (one $\sigma$ spin and two $\tau$ spins) are located at the boundaries.
The $\Z_2^{\sigma}$ symmetry acts weakly on the boundary $\sigma$ spins, and the spins in the ket and bra states are coupled.
Considering that the SWSSB of the $\Z_2^{\sigma}$ symmetry contributes a two-fold degeneracy as discussed above, the total GSD under OBC in this phase reads $\rm GSD_{ OBC} = 2 \text{ (SWSSB)} \times 2^3 \text{ (edge modes)} = 16$.
In contrast, the $\Z_2^{\sigma}$ (S) $\times$ $\Z_2^{\tau}$ (S) ASPT phase discussed above features four independent boundary spins (two $\sigma$ spins and two $\tau$ spins) as both $\Z_2$ symmetry are strong, leading to $2^4=16$ degenerate edge modes.
Table~\ref{Tab: GSD} summarizes the GSD under different boundary conditions and its connection to the topological and symmetry properties for each phase.

Finally, we turn to the critical lines and emergent triple points in the phase diagram.
In the upper half-plane ($\beta>0.5$), three critical lines define three types of gapless ASPT phases.
Each of these critical lines maps to a corresponding critical line in the lower half-plane ($1-\beta$) described by Ising criticality that breaks the associated symmetry.
A notable feature is the gapless ASPT criticality between the $\Z_2^{\sigma}$ (S) $\times$ $\Z_2^{\tau}$ (S) ASPT and the double ASPT phase.
This represents a new realization of the gapless ASPT, where both sides of the critical line exhibit the same decorated domain wall structure~\cite{Scaffidi2017} but with different symmetry properties.
This contrasts with the conventional emergence of gapless SPT phases in the transitions between a topologically nontrivial symmetric phase and a topologically trivial SSB phase in both closed~\cite{Scaffidi2017} and open systems~\cite{Guo2024C}.
Moreover, the intersection of three gapless ASPT phases (in the upper half-plane) further extends the concept of topological criticality to topological triple points, which presents an intriguing subject for investigation.
This topological triple point can be generated by decorating domain walls at the triple point between three symmetry-breaking critical lines (in the lower half-plane), inheriting features from both the symmetry-breaking and topological properties.
In summary, the interplay between symmetry and topology is much richer in an open quantum system originating from the distinct behaviors of strong and weak symmetries.

\emph{Conclusions and discussions.\textemdash}
In this work, we systematically investigate the phase diagram of a 1D open system with $\Z_2^{\sigma}\times \Z_2^{\tau}$ symmetry using the imaginary-Lindbladian formalism.
For the first time, we identify and construct the double ASPT phase, a new intrinsic mixed-state phase previously overlooked in the study of open quantum systems.
The concurrence of symmetry-breaking patterns and topological properties related to the same symmetry group in double ASPT phase suggests a distinct and richer phase diagram and classification in the context of open quantum systems.

By comparing the phase diagram of the pure-state system with that of the open system, we gain an intuitive understanding of how new physical phenomena emerge in the latter.
In particular, we observe three distinct realizations of $\Z_2^{\sigma}\times \Z_2^{\tau}$ gapless ASPT phases, as well as the appearance of topological triple points where these critical lines intersect.
Future investigations of the gapless ASPT phases and the topological triple points hold great promise.
A deeper exploration of the relationships and differences between these gapless ASPT phases, particularly from a topological perspective, is essential.
In particular, the topological triple point can be interpreted as a phase transition between two distinct gapless ASPT phases~\cite{Zhang2024}.
These insights are crucial for advancing our understanding of open-system phase transitions and are instrumental in constructing a comprehensive classification of gapless topological phases in open quantum systems.
 
\emph{Acknowledgments.\textemdash}
We thank Jian-Hao Zhang for helpful discussions. This work is supported by the National Natural Science Foundation of China (NSFC) (Grant No. 12475022, No. 12174214, and No. 92065205) and the Innovation Program for Quantum Science and Technology (Grant No. 2021ZD0302100).

\bibliography{ref}

\appendix
\onecolumngrid
\renewcommand{\theequation}{S\arabic{equation}} \setcounter{equation}{0}
\renewcommand{\thefigure}{S\arabic{figure}} \setcounter{figure}{0}

\newpage
\section*{Supplemental Material for `Strong-to-weak spontaneous symmetry breaking meets average symmetry protected topological order'}

In this supplemental material, we provide more details on the imaginary-time Lindbladian evolution formalism, construction of $\LL^I$ of SWSSB phase, calculation of various correlators, and more numerical results for phase diagrams. 
\section{Imaginary-time Lindbladian evolution}
In this section, we briefly review the recently proposed framework to define a gapped quantum phase in an open system as the finite-time steady state of the imaginary-time Lindbladian evolution~\cite{Guo2024C}.

The real-time evolution of an open system is usually characterized by the master equation~\cite{Breuer2007, Rivas2012}
\begin{align}
    \frac{\mathrm{d}\rho}{\mathrm{d}t} = -i[H, \rho]+\sum_k{\left[L_k\rho L_k^{\dagger}-\frac{1}{2}\{L_k^{\dagger}L_k, \rho\}\right]}
\end{align}
where $H$ is the Hamiltonian of the system, and $L_k$ is the jump operator to describe the coupling between system and environment.
To derive the imaginary-time evolution of the open system, we first consider the imaginary-time evolution for the entire system (system and environment) and trace out the ancillae, leading to the following expression~\cite{Guo2024C}
\begin{align}
    \frac{\mathrm{d}\rho}{\mathrm{d}\tau} = -\left\{H, \rho\right\}+ \sum_k \left[L_k \rho L_k^{\dagger} + \frac{1}{2}\{L_k^{\dagger}L_k, \rho\}\right] \equiv - \LL^I(\rho),
\end{align}
where $\LL^I$ is denoted as the imaginary-Liouville superoperator.
The steady state of $e^{-\LL^{I}\mathrm{d}\tau}$, if exists, is the ground state of $\LL^I$ (eigenvalue with the smallest real part), whose spectral gap determines the recursion time of converging to steady state $\tau\sim 1/\Delta^I$.
Using this formalism, one can construct the corresponding open system for typical mixed states belonging to different quantum phases and explore potential phase transitions along the path connecting two imaginary-Liouville superoperators through, e.g., linear interpolation.

\section{Construction of $\LL^I$ for SWSSB phase}
Here we construct the corresponding open system for the density matrix
\begin{align}
    \rho \equiv O_+^{\rm SWSSB}+O_-^{\rm SWSSB} = \left( \frac{1}{2^N} \prod_i \sigma_i^0 \right) + \left(\frac{1}{2^N} \prod_i\sigma_i^x \right).
\end{align}
We first review the heuristic method proposed in Ref.~\cite{Guo2024C} to construct the imaginary-Liouville superoperator.
Inspired by the fact that the classical mixture in a density matrix corresponds to superposition in the supervector space, we use the following two-step strategy: finding one term to project onto the manifold containing all possible configurations appearing in the superposition and designing another one to determine the superposition coefficients.
This approach can be used to construct the Hamiltonian for novel topological quantum states like 1D cluster state or 2D string-net states.

Let us consider the trivial symmetric mixed state
\begin{align}
    \rho = \frac{1}{2^N}\prod_i\left(\ket{\uparrow_i}\hspace{-0.5mm}\bra{\uparrow_i}+\ket{\downarrow_i}\hspace{-0.5mm}\bra{\downarrow_i}\right) = \frac{1}{2^{N}}\sum_{\{\sigma_i\}}\ket{\psi_{\{\sigma_i\}}}\hspace{-0.5mm}\bra{\psi_{\{\sigma_i\}}}.
\end{align}
We note that in each configuration constituting the density matrix, the spins in ket and bra are parallel, which can be achieved by the following ferromagnetic term
\begin{align}
     L_{i}^{[1]} = \sigma_i^z \Longrightarrow L_{i}^{[1]}\otimes L_{i}^{[1]*} = \sigma_i^z\otimes \sigma_i^z.
\end{align}
Meanwhile, the hopping terms between different components are chosen as
\begin{align}
    L_{i}^{[2]} = \sigma_i^x \Longrightarrow L_{i}^{[2]}\otimes L_{i}^{[2]*} = \sigma_i^x\otimes \sigma_i^x,
\end{align}
which connect two different configurations $\ket{\psi_{\{\sigma_i\}}}\hspace{-0.5mm}\bra{\psi_{\{\sigma_i\}}}$ and $\ket{\psi_{\{\sigma_i\}^{\prime}}}\hspace{-0.5mm}\bra{\psi_{\{\sigma_i\}^{\prime}}}$ that differ by the spins at site $i$.

As for the SWSSB case, the strategy is a little different.
The aim is to obtain a degenerate ground-state manifold spanned by $O_+$ and $O_-$, meaning that we need to stabilize both states but without coupling between them.
Notably, ket and bra spins on every site are completely parallel (antiparallel) for $\rho_+$ ($\rho_-$), or explicitly written as
\begin{align}
    O_+^{\rm SWSSB} &= \frac{1}{2^N}\prod_i\left(\ket{\uparrow_i}\hspace{-0.5mm}\bra{\uparrow_i}+\ket{\downarrow_i}\hspace{-0.5mm}\bra{\downarrow_i}\right),\\
    O_-^{\rm SWSSB} &= \frac{1}{2^N}\prod_i\left(\ket{\uparrow_i}\hspace{-0.5mm}\bra{\downarrow_i}+\ket{\downarrow_i}\hspace{-0.5mm}\bra{\uparrow_i}\right).
\end{align}
As an analog, in the GHZ state with an SSB order, each pair of adjacent sites is composed of $\ket{\uparrow_i \uparrow_{i+1}}$ or $\ket{\downarrow_i \downarrow_{i+1}}$ and corresponds to a ferromagnetic Hamiltonian $H=-\sum_i \sigma_i^z\sigma_{i+1}^z$.
Similarly, each pair of adjacent sites belongs to one of the following cases
\begin{align}
    \begin{tabular}{ll}
    ket and bra spins are parallel for both sites, &$\ket{\alpha_i}\hspace{-0.5mm}\bra{\alpha_i}\otimes \ket{\beta_{i+1}}\hspace{-0.5mm}\bra{\beta_{i+1}}$\\
    ket and bra spins are antiparallel for both sites, \quad\quad &$\ket{\alpha_i}\hspace{-0.5mm}\bra{\tilde{\alpha}_i}\otimes \ket{\beta_{i+1}}\hspace{-0.5mm}\bra{\tilde{\beta}_{i+1}}$
    \end{tabular}
\end{align}
This pattern can be realized by the following ferromagnetic term
\begin{align}
     L_{i}^{[1]} = \sigma_i^z\sigma_{i+1}^z \Longrightarrow L_{i}^{[1]}\otimes L_{i}^{[1]*} = \sigma_i^z\sigma_{i+1}^z\otimes \sigma_i^z\sigma_{i+1}^z.
\end{align}
Similarly, we use
\begin{align}
    L_{i}^{[2]} = \sigma_i^x \Longrightarrow L_{i}^{[2]}\otimes L_{i}^{[2]*} = \sigma_i^x\otimes \sigma_i^x,
\end{align}
to introduce hopping between different components within $O_+$ or $O_-$ since it does not change the relative direction between ket and bra of one site.
From the perspective of the Hilbert space dimension, a density matrix with $N$ sites can be viewed by a $2^{2N}$-dimensional supervector, requiring two independent jump operators at each site to remove redundancy.

\section{Various correlators: linear, R\'enyi-2, Wightman, and conditional mutual information}
The expectation value of an observable $O$ over a pure state $\ket{\psi}$ can be calculated as
\begin{align}
    \braket{O} = \braket{\psi|O|\psi}.
\end{align}
Specifically, we consider the correlation length $\xi$ defined from the connected correlation function
\begin{align}
\mathcal{C}(O, i, j) \equiv \braket{O_i O_j} - \braket{O_i}\braket{O_j}\sim \ e^{-|i-j|/\xi}.
\end{align}
These observables allow us to identify symmetry properties and long-range patterns that govern phase transitions in closed systems.

As for a density matrix in the open system, we consider three types of observable: the linear observable, defined as the conventional expectation value over a density matrix
\begin{align}
    \braket{O}\equiv \Tr[\rho O],
\end{align}
the R\'enyi-2 observable, expressed as the expectation value of the supervector $\sket{\rho}$~\cite{Guo2024A}
\begin{align}
    \braket{O}^{(2)}\equiv \frac{\sbraket{\rho|O\otimes O^*|\rho}}{\sbraket{\rho|\rho}} =  \frac{\Tr\left[\rho O \rho O^{\dagger}\right]}{\Tr\left[\rho^2\right]},
\end{align}
and the Wightman observable defined under canonical purification~\cite{Liu2024, Weinstein2024}
\begin{align}
    \braket{O}^{(\text{W})} \equiv \sbraket{\sqrt{\rho}|O\otimes O^{*}|\sqrt{\rho}} = \Tr{\left[\sqrt{\rho}O\sqrt{\rho}O\right]}.
\end{align}
Subsequently, their corresponding correlators can be constructed as follows, including the linear correlation function
\begin{align}
\mathcal{C}(O, i, j) \equiv \braket{O_i O_j} - \braket{O_i}\braket{O_j}\sim \ e^{-|i-j|/\xi},\label{Linear_corr}
\end{align}
the R\'enyi-2 correlation function
\begin{align}
    \mathcal{C}^{(2)}(O, i, j)\equiv \braket{O_i O_j}^{(2)} - \braket{O_i}^{(2)}\braket{O_j}^{(2)}\sim\ e^{-|i-j|/\xi_{2}},
\label{Renyi2}
\end{align}
and the Wightman correlation function
\begin{align}
    \mathcal{C}^{(\text{W})}(O, i, j)\equiv \braket{O_i O_j}^{(\text{W})} - \braket{O_i}^{(\text{W})}\braket{O_j}^{(\text{W})}\sim\ e^{-|i-j|/\xi_{\text{W}}},
\end{align}
which define the correlation lengths $\xi$, $\xi_2$, and $\xi_{\rm W}$, respectively.

In certain circumstances, the R\'enyi-2 correlator has distinct long-range behavior with the fidelity correlator, while the latter satisfies the stability theorem and is believed to be a universal indicator for SWSSB phases~\cite{Lessa2024B}.
However, the fidelity correlator is highly non-linear in terms of the density matrix and is thus intractable for numerical simulations.
Fortunately, it has been proposed that the Wightman correlator exhibits qualitatively equivalent long-range patterns to the fidelity correlator~\cite{Liu2024, Weinstein2024}.
Although still challenging to evaluate directly, this Wightman correlator can be viewed as a R\'enyi-1 correlator and estimated by analytical continuation from general R\'enyi-$n$ correlators (with even $n$) to $ n\rightarrow 1$~\cite{Liu2024}.
Specifically, a general R\'enyi-$n$ correlator can be defined as
\begin{align}
    \mathcal{C}^{(n)}(O, i, j)\equiv \braket{O_i O_j}^{(n)} - \braket{O_i}^{(n)}\braket{O_j}^{(n)}\sim\ e^{-|i-j|/\xi_{n}},
\end{align}
where
\begin{align}
    \braket{O}^{(n)}\equiv \frac{\Tr{\left[\rho^{\frac{n}{2}}O\rho^{\frac{n}{2}}O\right]}}{\Tr\left[\rho^n\right]}
\end{align}
denotes the R\'enyi-$n$ expectation value for a density matrix $\rho$.

The conditional mutual information (CMI), which is believed to characterize different phases of matter~\cite{Sang2025}, describes the mutual information of subsystems $A$ and $C$ conditioned on another subsystem $B$ as
\begin{align}
    I(A:C|B) = S(AB)+S(BC)-S(B)-S(ABC).
\end{align}
Here, $S(X)=-\Tr\left[\rho_X\ln \rho_X\right]$ refers to the von Neumann entropy of the reduced density matrix $\rho_X$.
The correlation length of CMI is known as Markov length $\xi_{\rm M}$, which is defined by the long-range behavior regarding the distance between subregions $A$ and $C$, or the width of the subregion $B$, i.e.,
\begin{align}
    I(A:C|B)\sim e^{-L_B/\xi_{\rm M}}.
\end{align}

\begin{figure}
    \centering
    \includegraphics[width=0.8\linewidth]{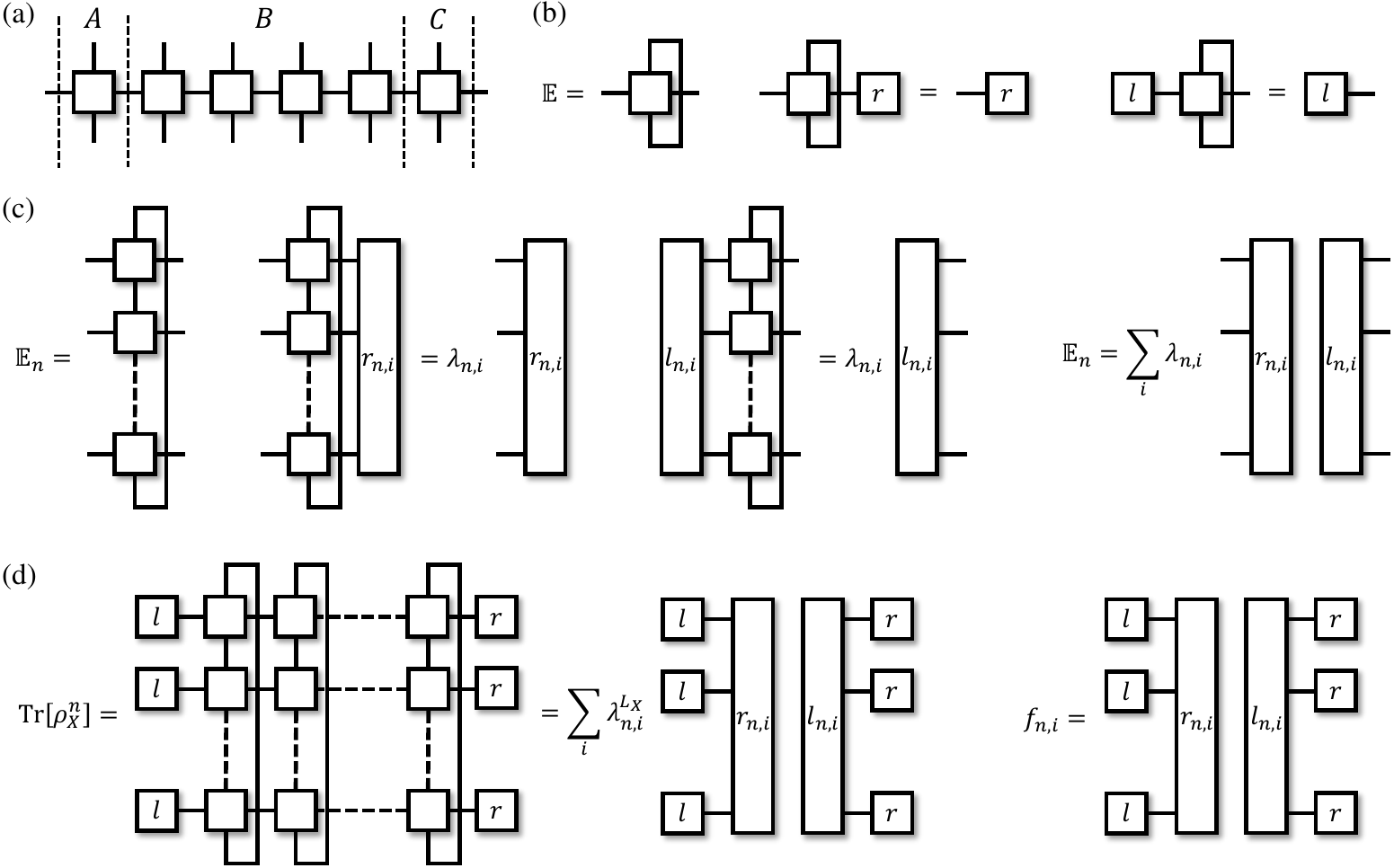}
    \caption{Calculation of CMI $I(A:C|B)$.
    (a) Division of $A$, $B$, and $C$ subregions.
    (b) Transfer matrix for linear expectation value with fixed points $l$ and $r$.
    (c) Transfer matrix for R\'enyi-$n$ expectation value and its spectral decomposition.
    (d) Calculation of R\'enyi-$n$ entropy by tensor contraction.}
    \label{Fig: TM}
\end{figure}

For 1D systems, we consider the division for $A$, $B$, and $C$ subregions as shown in Fig.~\ref{Fig: TM}(a), where $L_A$ and $L_C$ take constant values and we focus on the relation between $I(A:C|B)$ and $L_B$.
Similar to the Wightman correlator, we first calculate $I(A:C|B)^{(n)}$ defined by R\'enyi-$n$ entropy $S^{(n)}(X) = \frac{1}{1-n}\ln\Tr\left[\rho_X^n\right]$, then extrapolate to $n\rightarrow1$.
Suppose the density matrix $\rho$ does not exhibit long-range order in terms of the linear correlator, its transfer matrix $\mathbb{E}$ should possess a unique dominant eigenvector (or fixed point) $r$.
The corresponding eigenvalue can be properly normalized as $\lambda=1$ such that $\Tr[\rho]=1$, as shown in Fig.~\ref{Fig: TM}(b).
Correspondingly, the left fixed point of $\mathbb{E}$ is denoted as $l$.
Now we consider the transfer matrix $\mathbb{E}_n$ shown in Fig.~\ref{Fig: TM}(c) for calculating
\begin{align}
    \Tr[\rho_X^n] = l^{\otimes n} \cdot \mathbb{E}_n^{L_X} \cdot r^{\otimes n},
\end{align}
where $X$ is a continuous region with a finite length $L$.
Therefore, by preserving the largest two eigenvalues (in the sense of magnitude) $\lambda_{n, 0}$ and $\lambda_{n, 1}$, we can simplify the expression for $S^{(n)}(X)$ as
\begin{align}
\begin{aligned}
    S^{(n)}(X) &= \frac{1}{1-n}\ln \Tr[\rho_X^n]\\
    &\approx \frac{1}{1-n} \ln\left[\lambda_{n, 0}^{L_X} \left(l^{\otimes n} \cdot  r_{n, 0}\right) \left(l_{n, 0}\cdot r^{\otimes n}\right) + \lambda_{n, 1}^{L_X} \left(l^{\otimes n} \cdot  r_{n, 1}\right) \left(l_{n, 1}\cdot r^{\otimes n}\right)\right] \\
    & = \frac{1}{1-n} \ln \left\{f_{n, 0} \lambda_{n, 0}^{L_X} \left[1+\frac{f_{n, 1}}{f_{n, 0}} \left(\frac{\lambda_{n, 1}}{\lambda_{n, 0}}\right)^{L_X}\right]\right\}\\
    & = \frac{1}{1-n} \left\{\ln f_{n, 0} +L_X\ln \lambda_{n, 0}+ \ln \left[1+\frac{f_{n, 1}}{f_{n, 0}} \left(\frac{\lambda_{n, 1}}{\lambda_{n, 0}}\right)^{L_X}\right]\right\}\\
    & \approx \frac{1}{1-n} \left[\ln f_{n, 0} +L_X\ln \lambda_{n, 0}+ \frac{f_{n, 1}}{f_{n, 0}} \left(\frac{\lambda_{n, 1}}{\lambda_{n, 0}}\right)^{L_X}\right].
\end{aligned}
\end{align}
In the third equation, we denote constants $f_{n, i} \equiv \left(l^{\otimes n} \cdot  r_{n, i}\right) \left(l_{n, i}\cdot r^{\otimes n}\right)$ for $i=0, 1$ as shown in Fig.~\ref{Fig: TM}(d), while in the last approximation, we adopt the property $\ln (1+x)\approx x$ for $x\ll 1$.
Therefore, for the division shown in Fig.~\ref{Fig: TM}(a), CMI for R\'enyi-$n$ entropy can be derived as
\begin{align}
\begin{aligned}
    I^{(n)}(A:C|B) &= \frac{1}{1-n} \frac{f_{n, 1}}{f_{n, 0}}\left[\left(\frac{\lambda_{n, 1}}{\lambda_{n, 0}}\right)^{L_A+L_B}+\left(\frac{\lambda_{n, 1}}{\lambda_{n, 0}}\right)^{L_B+L_C}-\left(\frac{\lambda_{n, 1}}{\lambda_{n, 0}}\right)^{L_B}-\left(\frac{\lambda_{n, 1}}{\lambda_{n, 0}}\right)^{L_A+L_B+L_C}\right]\\
    &= \frac{1}{n-1} \frac{f_{n, 1}}{f_{n, 0}}\left(\frac{\lambda_{n, 1}}{\lambda_{n, 0}}\right)^{L_B}\left[1-\left(\frac{\lambda_{n, 1}}{\lambda_{n, 0}}\right)^{L_A}\right]\left[1-\left(\frac{\lambda_{n, 1}}{\lambda_{n, 0}}\right)^{L_C}\right]\propto e^{-L_B/\xi_n},
\end{aligned}
\end{align}
for fixed $L_A$ and $L_C$ and non-degenerate $\lambda_{n, 1}\neq \lambda_{n, 0}$.
Here, $\xi_n\equiv - \left({\ln\left|\frac{\lambda_{n, 1}}{\lambda_{n, 0}}\right|}\right)^{-1}$ is nothing but the correlation length for the R\'enyi-$n$ correlation function.
On the other hand, if the spectrum of $\mathbb{E}_n$ has degeneracy with $\lambda_{n, 1}= \lambda_{n, 0}$, e.g., belonging to the SWSSB phase, the contribution from $\lambda_{n, 1}$ will be eliminated and the third largest eigenvalue $\lambda_{n, 2}$ will become the dominant term for the above equation in the long-range limit.
For non-injective MPO with the direct-sum structure $\rho=O_+ + O_-$, $\lambda_{n, 2}$ of $\rho$ is equivalent to $\lambda_{n, 1}$ of single $O_+$.
This implies that CMI of $\rho$ is also short-range correlated even in the SWSSB phase and shares the same correlation length with $O_+$.

\begin{figure}
    \centering
    \includegraphics[width=0.85\linewidth]{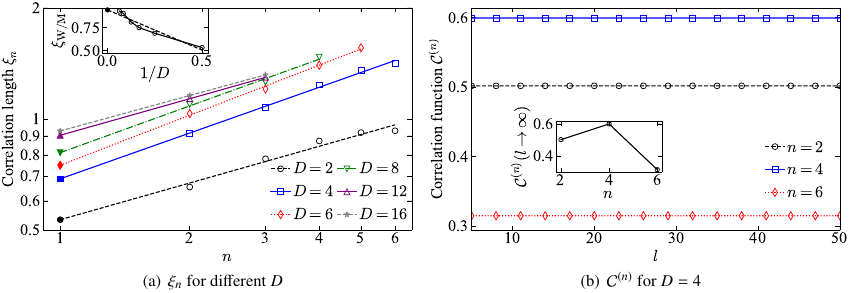}
    \caption{Extrapolation of $\xi_{\rm{W/M}}$ for $\beta=0.6$ of the $\Z_2$ SWSSB model.
    (a) Extrapolation of $\xi_n$ to $n\rightarrow 1$ and $1/D\rightarrow 0$. 
    (b) $\mathcal{C}^{(n)}$ for different $n$ exhibit a non-monotonic behavior.}
    \label{Fig: Extrapolate}
\end{figure}

In summary, both the Wightman correlation length $\xi_{\rm W}$ and Markov length $\xi_{\rm M}$ can be estimated by extrapolation of R\'enyi-$n$ correlation length $\xi_{n}$ to $n\rightarrow 1$.
However, the calculation of $\xi_n$ still poses additional complexity for large $n$ and bond dimension $D$ of MPO, where the dimension of transfer matrix $\mathbb{E}_n$ is $D^n \times D^n$.
Therefore, we first extrapolate $\xi_n$ to $n \rightarrow 1$ for different $D$, then extrapolate $\xi_{\rm W/M}$ to $D\rightarrow \infty$.
Fig.~\ref{Fig: Extrapolate}(a) shows a typical implementation of this process for $\beta=0.6$ of the $\Z_2$ SWSSB model studied in Fig.~1.
In particular, $\xi_n$ does not show parity dependencies, validating a simultaneous extrapolation of $\xi_{\rm W}$ and $\xi_{\rm M}$.
On the other hand, direct fitting of $\mathcal{C}^{(n)}$ is difficult since the concrete values of correlators are not universal, as demonstrated by the non-monotonic behavior of $\mathcal{C}^{(n)}$ regarding different $n$ in Fig.~\ref{Fig: Extrapolate}(b).
Moreover, we demonstrate that CMI of $\rho=O_++O_-$, i.e., with the symmetry recovered, is also short-range correlated in both trivial and SWSSB phases by comparing Fig.~\ref{Fig: SWSSB}(b-c).
Specifically, $I^{(2)}$ of $\rho$ has the same correlation length as $\mathcal{C}^{(2)}$ of simple $O _+$ (which we refer to as $\mathcal{C}^{(2)}_+$) as shown in Fig.~\ref{Fig: SWSSB}(c), verifying our previous analytical results.
Therefore, $I^{(2)}$ has different long-range patterns with $\mathcal{C}^{(2)}$ or $\mathcal{C}^{(\rm{W})}$ and is not able to serve as an indicator for the existence of SWSSB.
Nevertheless, the divergence of $\xi_{\rm M}$ still demonstrates the occurrence of phase transition, consistent with previous studies~\cite{Sang2025}.
The behaviors of different correlation functions are summarized in Table~\ref{Tab: Corr} for both $\rho$ and $O_+$.

\begin{figure}
    \centering
    \includegraphics[width=0.8\linewidth]{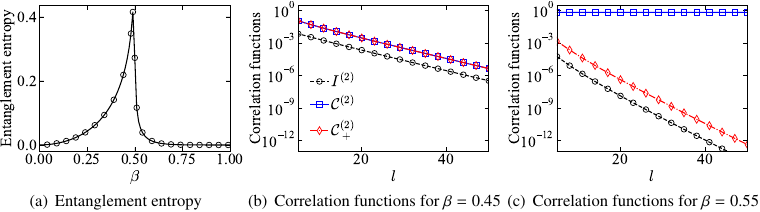}
    \caption{Additional numerical results for $\Z_2$ SWSSB model.
    (a) Entanglement entropy of the supervector $\sket{O_+}$.
    (b) Various correlation functions in the trivial phase with $\beta=0.45$.
    (c) Various correlation functions in the SWSSB phase with $\beta=0.55$.}
    \label{Fig: SWSSB}
\end{figure}

\begin{table}[H]
\centering
\caption{Long-range behaviors of various correlators for $\rho$ and $O_+$ in the SWSSB phase.}
\begin{tabular}{c|c|c|c|c}\hline\hline
    State & Linear $\mathcal{C}$ & R\'enyi-$2$ $\mathcal{C}^{(2)}$ & Wightman $\mathcal{C}^{(\rm W)}$ & Conditional mutual information $I^{(2)}$ \\\hline
     $O_+$ & $e^{-l/\xi}$ & $e^{-l/\xi_2}$ & $e^{-l/\xi_{\rm W}}$ & $e^{-l/\xi_2}$\\
     $\rho=O_++O_-$ & $e^{-l/\xi}$ & $O(1)$ & $O(1)$ & $e^{-l/\xi_2}$\\\hline\hline
\end{tabular}
\label{Tab: Corr}
\end{table}

\section{Imaginary Liouville spectrum, Correlation lengths, symmetry indicators, and string order}
In this section, we present additional numerical results to compare the phase diagram of closed and open systems with $\Z_2^{\sigma}\times \Z_2^{\tau}$ symmetry with various physical properties, including imaginary Liouville spectrum, correlation lengths, symmetry indicators, and the string order parameter.

\begin{figure}
    \centering
    \includegraphics[width=0.8\linewidth]{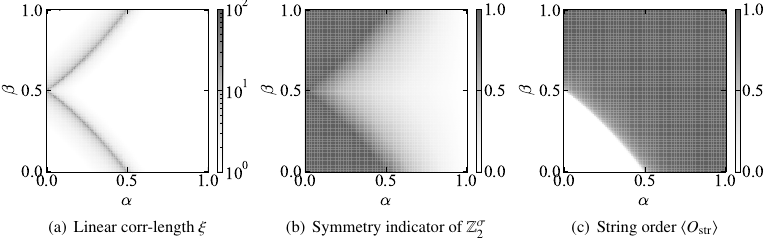}
    \caption{Physical properties for the closed system.
    (a) Linear correlation length $\xi$.
    (b) Symmetry indicator of $\Z_2^{\sigma}$.
    (c) String order parameter $\braket{O_{\rm{str}}}$.}
    \label{Fig: closed}
\end{figure}

For the closed system, we plot the correlation length $\xi$ in Fig.~\ref{Fig: closed}(a), revealing three quantum phases in the phase diagram divided by two critical lines with divergent $\xi$.
Meanwhile, the symmetry indicator for $\Z_2^{\sigma}$ symmetry
\begin{align}
    |\braket{K}| \equiv \left|\braket{\prod_i \sigma_i^{x}}\right|
\end{align}
in Fig.~\ref{Fig: closed}(b) demonstrates the occurrence of SSB for $\Z_2^{\sigma}$ symmetry, resulting in a $\Z_2^{\sigma}$ SSB $\times$ $\Z_2^{\tau}$ trivial phase in the intermediate region.
Finally, the string order to detect the decorated domain wall structure
\begin{align}
    \braket{O_{\rm str}} \equiv \lim_{|i-j|\rightarrow \infty}\braket{\sigma_{i}^z\tau_{i+1/2}^{x}\tau_{i+3/2}^x\cdots \tau_{j-3/2}^x\tau_{j-1/2}^x\sigma_{j}^z}
\end{align}
in Fig.~\ref{Fig: closed}(c) suggests a nontrivial $\Z_2^{\sigma}\times \Z_2^{\tau}$ SPT phase in the top region.
Notably, both SPT and SSB will lead to a long-range string order, underscoring the importance of combining different indicators to verify which quantum phase each region represents.
We note that all these observables are calculated for quantum states under a small perturbing field $H_i^{\prime} = h\sigma_i^z$ with $h=10^{-3}$ for numerical stability.

\begin{figure}
    \centering
    \includegraphics[width=0.8\linewidth]{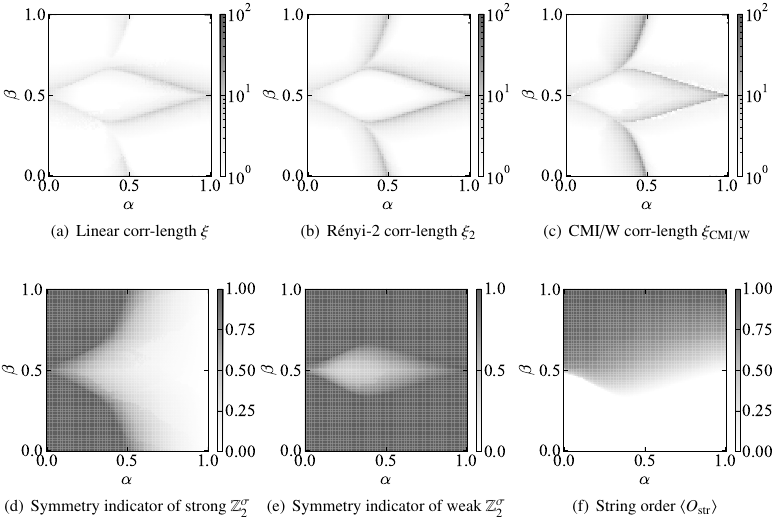}
    \caption{Physical properties for the open system.
    (a) Linear correlation length $\xi$.
    (b) R\'enyi-2 correlation length $\xi_2$.
    (c) Wightman/Markov correlation lengths $\xi_{\rm W/M}$.
    (d) Symmetry indicator of strong $\Z_2^{\sigma}$.
    (e) Symmetry indicator of weak $\Z_2^{\sigma}$.
    (f) String order parameter $\braket{O_{\rm{str}}}$.}
    \label{Fig: open}
\end{figure}

As for the open system, all correlation lengths (of $O_+$) exhibit the same divergent properties in Fig.~\ref{Fig: open}(a-c), dividing the parameter space into five quantum phases.
Similar to the pure-state case, we subsequently evaluate the symmetry properties for $\Z_2^{\sigma}$ symmetry.
The indicators for determining whether a mixed state satisfies the strong and weak $Z_2^{\sigma}$ symmetry are~\cite{Guo2024C}
\begin{align}
    |\braket{K}| &= \left|\Tr\left[\rho\prod_i\sigma_{i}^x\right]\right|,\\
    \braket{K}^{(2)} &= \frac{\Tr\left[\rho \left(\prod_i{\sigma_i^x}\right) \rho\left(\prod_i{\sigma_i^x}\right)\right]}{\Tr\left[\rho^2\right]},
\end{align}
respectively.
The results in Fig.~\ref{Fig: open}(c-d) indicate the SWSSB of $\Z_2^{\sigma}$ symmetry in the top right and bottom right regions inherited from our corner imaginary-Liouville superoperators $\LL^I$ in Fig.~2(b).
As for the strong and weak $\Z_2^{\tau}$ symmetry, we can define the corresponding symmetry indicators similarly, where the results show that the entire phase diagram preserves $\Z_2^{\tau}$ (S) symmetry.
Moreover, we observe the emergence of an SNSSB phase in the intermediate region, which is not explicitly constructed in our model.
Finally, the string order
\begin{align}
    \braket{O_{\rm str}} \equiv \lim_{|i-j|\rightarrow \infty}\braket{\sigma_{i}^z\tau_{i+1/2}^{x}\tau_{i+3/2}^x\cdots \tau_{j-3/2}^x\tau_{j-1/2}^x\sigma_{j}^z}
\end{align}
shown in Fig.~\ref{Fig: open}(e) suggests that both the ASPT phase protected by $\Z_2^{\sigma}$ (S) $\times$ $\Z_2^{\tau}$ (S) and the one protected by $\Z_2^{\sigma}$ (W) $\times$ $\Z_2^{\tau}$ (S) ($\rho_+$ in the double ASPT phase belongs to a simple ASPT after SWSSB of $\Z_2^{\sigma}$) exhibit long-range string order due to their similar decorated domain wall structures.
Therefore, to distinguish between these two phases, joint consideration of the symmetry indicators in Fig.~\ref{Fig: open}(c-d) is necessary.

\begin{figure}
    \centering
    \includegraphics[width=\linewidth]{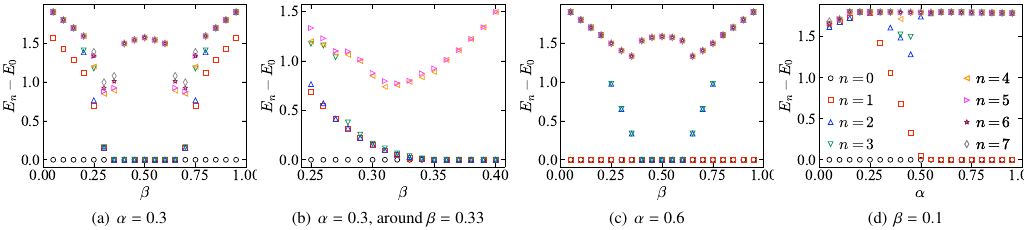}
    \caption{Imaginary-Liouville spectrum (the smallest eight eigenvalues) along several lines in the phase diagram with $N=10$ under PBC.
    (a) $\alpha=0.3$.
    (b) $\alpha=0.3$ and around the phase transition at $\beta=0.33$.
    (c) $\alpha=0.6$.
    (d) $\beta=0.1$.}
    \label{Fig: Spectrum}
\end{figure}

Finally, we plot the imaginary-Liouville spectrum along the lines of $\alpha=0.3$, $\alpha=0.6$, and $\beta=0.1$ in Fig.~\ref{Fig: Spectrum} for $N=10$ under PBC to demonstrate the properties of imaginary Liouville gap and steady-state degeneracy in each phase, especially near phase transitions.
In short, all phase transitions show similar gap-closing patterns at the transition points, while phases with different symmetries to be broken have different degeneracies.
For instance, the transition between the trivial phase and the SNSSB phase demonstrated along the $\alpha=0.3$ line in Fig.~\ref{Fig: Spectrum}(a-b) shows the merging of the ground state and the lowest three excited states when transitioning to the SNSSB phase characterized by a four-fold degeneracy.
In contrast, Fig.~\ref{Fig: Spectrum}(c) reveals another type of gap closing that is characterized by the gap between the steady-state manifold (that is two-fold degenerate) and the higher excited states when transitioning from the SWSSB phase to the SNSSB phase.
As for the upper half of the phase diagram, though the critical lines are topologically nontrivial and belong to gapless ASPT phases, they share similar spectrum properties due to the DW duality in our model.

\end{document}